\documentclass[pra,twocolumn]{revtex4}
\usepackage{graphicx}

\begin{document}
\title{Nonmaximally-entangled-state quantum photolithography}
\author{ Yan-Hui Wang and Le-Man Kuang\footnote{Corresponding author.}\footnote{
Email: lmkuang@hunnu.edu.cn}}
\address{Department of Physics, Human Normal University, Changsha 410081, China\ }

\begin{abstract}
Many previous works on quantum photolithography are based on
maximally-entangled states (MES).  In this paper, we generalize
the MES quantum photolithography to the case where two light beams
share a $N$-photon nonmaximally-entangled state. we investigate
the correlations between quantum entanglement and quantum
photolithography.  It is shown that for nonlocal entanglement
between the two light beams the amplitude of the deposition rate
can be changed through varying the degree of entanglement
described by an entanglement angle while the resolution remains
unchanged, and found that for local entanglement between the two
light beams the effective Rayleigh resolution of quantum
photolithography can be resonantly enhanced.

\vspace{0.5cm}
 \noindent PACS number(s): 42.50.Dv, 42.25.Hz,
85.40.Hp

\end{abstract}

\maketitle

\section{Introduction}
Recently, much attention has been paid to quantum photolithography
\cite{r1,r2,r3,r3'} due to the possibility of beating the
classical Rayleigh diffraction limit through using on quantum
entanglement between two used light beams. Optical lithography is
a widely used printing method which has been the primary  tool of
the semiconductor industry for transferring circuit images onto
substrates to produce smaller and smaller processors. In this
process, light is used to etch a substrate and the exposed or
unexposed areas on the substrate define  patterns. The resolution
of images transferred by using classical light beams is restricted
to the Rayleigh diffraction limit $\lambda/4$, $\lambda$  being
the wavelength of the light, hence one can achieve a resolution
only comparable to the wave length of the light used in classical
optical lithography \cite{r4,r5,r6}. In Ref. \cite{r3}, Agedi and
coworkers introduced a procedure called quantum lithography in one
dimension that predicts an increase in resolution beyond the
diffraction limit due to quantum entanglement between two light
beams, and demonstrate a quantum lithography method to improve the
resolution by a factor $N$ in contrast to classical one, using
$N$-photon maximal entangled state (MES) \cite{r7}. The Maryland
group \cite{r2} completed  a proof-of-principle experimental
demonstration of quantum lithography by using two-photon entangled
state generated via a specially designed spontaneous parametric
down-conversion. The increase in resolution makes quantum
lithography a potentially useful tool to produce smaller computer
chips in nanotechnology. Then the one-dimensional quantum
lithography method was generalized to the two-dimensional case
\cite{r8} and entangled binomial states \cite{r9}. Since the
number of elements writable on a surface scales inverse
quadratically with the minimum feature dimension, this improvement
is an important advance. The purpose of the present paper is to
generalize the MES quantum lithography to the case where two light
beams share a $N$-photon nonmaximally-entangled state (NMES), and
investigate properties of the deposition rate for general patterns
in one dimension.

This paper is organized as follows. In  Sec. II, we introduce the
NMES quantum lithography method after briefly recalling  the MES
quantum lithography method. In Sec. III we discuss pattern
engineering in one dimension. Finally, we summarize our results in
Sec. IV.

\section{NMES quantum photolithography}

The quantum photolithographic process is based on the multi-photon
absorption process on a substrate.  The character parameter of the
optical lithography is the minimal resolvable feature size which,
according to Rayleigh criterion, occurs at a spacing corresponding
to the distance between an intensity maximum and an adjacent
intensity minimum \cite{r9}, or optical resolution which may
denote the minimal distance between two nearby points that can
still be resolved with microscopy or the minimal distance
separating two points that are printed using lithography. For a
$N$-photon state $|\psi_N\rangle$ the resolution is determined by
the absorption rate at the imagine surface which is proportional
the expectation value of the dosing operator
\begin{equation}
\label{1}
\Delta_N(\phi)=\langle\psi_N|\hat{\delta}_N|\psi_N\rangle,
\end{equation}
where the dosing operator is defined by
\begin{equation}
\label{2}
\hat{\delta}_N=\frac{(\hat{e}^{\dagger})^N\hat{e}^N}{N!},
\end{equation}
with the superposition mode operator
$\hat{e}=(\hat{a}+\hat{b})/\sqrt{2}$ and its adjoint
$\hat{e}^{\dagger}=(\hat{a}^{\dagger}+\hat{b}^{\dagger})/\sqrt{2}$.
If a substrate is exposed for a time $t$ to the light source, the
exposure function $P(\phi)=\Delta_Nt$ gives an exposure pattern.


Before going into the NMES lithography, let us briefly review the
MES  quantum lithography presented in Ref \cite{r3}. Consider  two
counterpropagating light beams $a$ and $b$ cross each other at the
surface of a photosensitive substrate. They have a relative phase
difference $\phi=kx/2$ with the optical wave number
$k=2\pi/\lambda$ and $x$ is the lateral dimension on the substrate
to describe the position where the two beams meet. For the
$N$-photon maximally-entangled number state of the two light beams
\begin{equation}
\label{3}
|\psi_N\rangle_{ab}=\frac{1}{\sqrt{2}}(|N,0\rangle_{ab}+e^{iN\phi}|0,N\rangle_{ab}),
\end{equation}
from Eqs. (\ref{1}) and (\ref{2}) one can get the  following
deposition rate
\begin{equation}
\label{4} \Delta_N(\phi)=\frac{1}{2^N}(1+\cos N\phi),
\end{equation}
which indicates that a $2\pi$ shift of $\phi$ will displace $N$
times. Hence, the $N$-photon MES (\ref{1}) produces an effective
Rayleigh resolution given by
\begin{equation}
\label{5} \Delta x=\frac{\lambda}{4N},
\end{equation}
which increases the resolution by a factor $N$ in contrast to the
classical diffraction limit $\Delta x=\lambda/4$.

Now we consider the quantum lithography with a $N$-photon
 NMES given by
\begin{equation}
\label{6}
|\psi_N\rangle_{ab}=\cos\gamma|N,0\rangle_{ab}+e^{iN\phi}\sin\gamma|N,0\rangle_{ab},
\end{equation}
where $\gamma$ measures the entanglement of the state, it changes
from $0$ (no entanglement) to $\pi/4$ (maximal entanglement).

For the $N$-photon NMES  (\ref{6}), from Eqs. (\ref{1}) and
(\ref{2}) we obtain  the following deposition rate
\begin{equation}
\label{7}
\Delta_N(\gamma,\phi)=\frac{1}{2^N}\left[1+\sin(2\gamma)\cos(N\phi)\right].
\end{equation}

In what follows, we consider the influence of quantum entanglement
for the two cases of nonlocal and local entanglement,
respectively.

{\bf Case 1. }{\it Nonlocal entanglement}. In this case, the
entangling angle of the two light beams  $\gamma$ is independent
of the position where the two beams meet, then $\gamma$ is
irrelevant  to the phase difference $\phi$ in the NMES (\ref{6}).
From the expression of the deposition rate (\ref{7}) we can see
that the amplitude of the deposition function increases with the
degree of entanglement. Hence one can manipulate and control the
amplitude of the deposition rate through varying the quantum
entanglement between the two light beams.

{\bf Case 2. }{\it Local entanglement}. In this case, the
entangling angle of the two light beams  $\gamma$ is dependent of
the position where the two beams meet, then $\gamma$ is relevant
to the phase difference $\phi$ in the NMES (\ref{6}). From Eq.
(\ref{7}) we can see that this type of phase relevance, which is
induced by the local entanglement, affects the period of the
deposition function. It is the period of the deposition function
that  determines the resolution of the  quantum photolithography.
Hence, the local entanglement between the two light beams can
improve the resolution. In particular,  if we suppose that the
entanglement angle changes with variation of the phase $\phi$
according to the resonant relation $2\gamma=kN\phi$ with $k$ being
positive integers, then the deposition function (\ref{7}) becomes
\begin{equation}
\label{8} \Delta_N(\phi)=\frac{1}{2^{N+1}}\left[2+\sin(k+1)N\phi +
\sin(k-1)N\phi\right],
\end{equation}
which leads to the following resolution
\begin{equation}
\label{9} \Delta x=\frac{1}{4(k+1)N}.
\end{equation}
which indicates that resonant change of  the entanglement angle of
the $N$-photon NMES  enhances the resolution by a factor in
contrast to the case of the $N$-photon MES.   Therefore, one can
improve the resolution through resonantly changing the
entanglement angle of the $N$-photon NMES.

In optical lithography, one usually wishes that produced patterns
are uniform. However, the second and third terms on the right-hand
side of Eq. (\ref{8}) have different periods with respect to the
phase $\phi$. In general superpositions of periodic functions with
different modulation periods produces so-called
collapse-and-revival phenomena which appear in laser-atom
interactions \cite{r11,r12}. These collapses and revivals produce
are non-uniform patterns. This non-uniform pattern problems can be
removed through choosing specific values of $k$ in Eq. (\ref{8}).
If we set $k=1$, the deposition function becomes
\begin{equation}
\label{8'}
\Delta_N(\phi)=\frac{1}{2^{N+1}}\left[2+\sin(2N\phi)\right],
\end{equation}
which leads to the following resolution
\begin{equation}
\label{9'} \Delta x=\frac{1}{8N},
\end{equation}
which implies that uniform patterns can be produced and the
resolution is doubled with respect to the MES quantum lithography.

\section{Pattern engineering in one dimension}

In this section,  we show how to manipulate and control patterns
in one dimension using the $N$-photon NMES and superposition
principle in quantum mechanics. In order to engineer arbitrary
one-dimensional paterns, we consider a more general $N$-photon
NMES
\begin{eqnarray}
\label{10}
|\psi_{Nm}\rangle_{ab}&=&e^{im\phi}\cos\gamma|N-m,m\rangle_{ab} \nonumber \\
& &+ e^{i(N-m)\phi}e^{i\theta_m} \sin\gamma|m,N-m\rangle_{ab},
\end{eqnarray}
which is an extension of the $N$-photon NMES given by Eq.
(\ref{6}) and reduces to  Eq. (\ref{6}) when $m=0$ and
$\theta_m=0$.  In order to obtain the deposition rate
corresponding to the $N$-photon NMES (\ref{10}), we have to
calculate the matrix elements of the dosing operator
\begin{equation}
\label{11}
\Delta^{Nm'}_{Nm}=\langle\psi_{Nm}|\hat{\delta}_N|\psi_{Nm'}\rangle,
\end{equation}
which is given by
\begin{eqnarray}
\label{12} \Delta^{Nm'}_{Nm}(\gamma,\phi)&=&\frac{1}{2^{N}}
\sqrt{C^m_NC^{m'}_N}\left[\cos^2\gamma e^{i(m'-m)\phi} \right.\nonumber \\
& & \left.+ \sin^2\gamma e^{-i(m'-m)\phi}
e^{i(\theta_{m'}-\theta_m)}\right. \nonumber \\
& & \left. + \frac{1}{2}\sin2\gamma [e^{i(N-m-m')\phi}
e^{i\theta_{m'}} \right. \nonumber \\
& & \left.+  e^{-i(N-m-m')\phi} e^{-i\theta_m}] \right],
\end{eqnarray}
where we have used the symbol $C^m_N=N!/(N-m)!m!$.

Then the expectation value of the dosing operator with respect to
the general $N$-photon NMES   (\ref{10}) can be obtained  by
taking the condition  $m=m'$ in Eq. (\ref{10})
\begin{equation}
\label{13}
\Delta_{Nm}(\gamma,\phi)=\frac{1}{2^{N}}C^m_N\{1+\sin(2\gamma)\cos[(N-2m)\phi+\theta_m]\}
\end{equation}

In particular, when $\gamma=\pi/2$, the deposition rate (\ref{13})
becomes
\begin{equation}
\label{14}
\Delta_{Nm}(\pi/2,\phi)=\frac{1}{2^{N}}C^m_N\{1+\cos[(N-2m)\phi+\theta_m]\}.
\end{equation}
Thus, we recover the deposition rate for the case of the
$N$-photon MES \cite{r5}.

One important point to be emphasized is that based on the
$N$-photon NMES   (\ref{10})  one can design various types of
patterns on a substrate through building various different kinds
of superposition states of the  $N$-photon NMES (\ref{10}). We
here consider the pseudo-Fourier method  \cite{r7} where one uses
a general superposition state with different photon numbers $n$
and a fixed distribution $m$ over two modes
\begin{equation}
\label{15} |\Phi_{N}\rangle=\sum_{n=0}^{N}C_{n}|\psi_{nm}\rangle,
\end{equation}
where the basic state $|\psi_{nm}\rangle$ is defined by (\ref{10})
through replacing $N$ with $n$,it is a $n$-photon NMES, and $C_n$
is an arbitrary coefficients. Hence, the state $|\psi_{nm}\rangle$
is a superposition state of states with different total photon
numbers in each branch.

Since
$\left[(\hat{a}+\hat{b})^N|\Phi_{N}\rangle\right]^{\dagger}=\langle\Phi_{N}|(\hat{a}^{\dagger}+\hat{b}^{\dagger})^N$,
branches with different photon numbers $n$ and $n'$ do not exhibit
interference when we calculate  the $N$-photon deposition rate of
the superposition state (\ref{15}) which  can be written as
\begin{eqnarray}
\label{16} \Delta_m=\sum_{n=0}^{N}|C_{n}|^2
\langle\psi_{nm}|\hat{\delta}_N|\psi_{nm}\rangle.
\end{eqnarray}

\begin{figure}[htb]
\begin{center}
\includegraphics[width=8.5cm,height=6.8cm,angle=0]{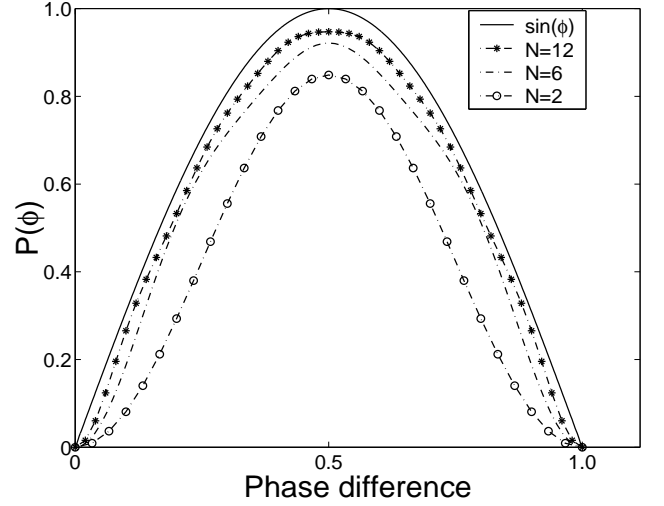}
\end{center}
\vskip 0.2cm \caption{The simulation of the $|\sin\phi|$-type
pattern  on the substrate resulting from a superposition of states
with $N=2,6$ and $12$, respectively. The solid curve is the
simulated pattern of the test function $|\sin\phi|$. Here the
phase difference is in units of $\pi$.} \label{kw}
\end{figure}

Making use of  Eqs. (\ref{10}) to (\ref{13}), from (\ref{16}) we
obtain
\begin{eqnarray}
\label{17}
\Delta_m(\gamma,\phi)=\sum_{n=0}^{N}|C_{n}|^2\Delta_{nm}(\gamma,\phi),
\end{eqnarray}
where $\Delta_{nm}(\gamma,\phi)$ is given by Eq. (\ref{13}) with
replacing $N$ by $n$. Eq. (\ref{17}) implies that the deposition
rate of the superposition state (\ref{15}) depends on only the
module of the superposition  coefficients $C_n$, and it is
independent of the phase of $C_n$.

 Using the expression of exposure pattern $P(\phi)=\Delta_mt$, from Eqs. (\ref{13}) and (\ref{17}) we
 find that
\begin{equation}
\label{18}
P(\phi)=\frac{t}{2^{N}}\sum_{n=0}^{N}C^m_N|C_n|^2\{1+\cos[(n-2m)\phi+\theta_m]\sin(2\gamma)\},
\end{equation}
which indicates that the exposure pattern induced by the
superposition state (\ref{15}) is determined by the module of the
superposition coefficients $|C_n|$, the entanglement angle
$\gamma$, and the relative phase $\theta_n$.

From  Eq. (\ref{18}) it is easy to understand the role of quantum
entanglement in quantum photolithography. In fact,  from  Eq.
(\ref{18})  we can see that varying of the entanglement angle
$\gamma$ is equivalent to rescaling time parameter $t$ or/and the
module of the superposition coefficient $|C_n|$ for the case of
nonlocal entanglement where the entanglement angle $\gamma$ is
independent of the associated phase difference $\phi$. In
particular, when $\gamma=\pi/4$, Eq. (\ref{18}) reduces to the
expression of exposure pattern of the MES case \cite{r7}
\begin{equation}
\label{19}
P(\phi)=\frac{t}{2^{N}}\sum_{n=0}^{N}C^m_N|C_n|^2\{1+\cos[(n-2m)\phi+\theta_n]\}.
\end{equation}

For the case of nonlocal entanglement the expression of exposure
pattern (\ref{18}) can be written as the sum of a general uniform
background exposure of the substrate and a standard truncated
Fourier series
\begin{equation}
\label{20} P(\phi)=Qt + t\sum_{n=0}^{N}(a_n\cos n\phi + b_n\sin
n\phi),
\end{equation}
where $Q$ is the uniform background penalty exposure rate
\begin{equation}
\label{21} Q=\sum^N_{n=0}|C_n|^2,
\end{equation}
and  the expanding coefficients are determined by the module of
the superposition coefficients $|C_n|$, the entanglement angle
$\gamma$, and the relative phase $\theta_n$ with the following
expressions
\begin{equation}
\label{22} a_n=|C_n|^2\sin(2\gamma)\cos \theta_n, \hspace{0.3cm}
b_n=|C_n|^2\sin(2\gamma)\sin\theta_n.
\end{equation}

From Eqs. (\ref{20}), (\ref{21}), and (\ref{22}) we can see that
quantum entanglement between two light beams dos not change the
background penalty exposure rate but it controls the amplitudes of
all Fourier components.

It is well known that any sufficiently well-behaved periodic
function can be written as an infinite Fourier series. However,
when we create patterns with the pseudo-Fourier lithography
method, we do not have access to every component of the Fourier
expansion, since this would involve an infinite number of photons
. This means that we can only employ truncated Fourier series, and
these can merely approximate arbitrary patterns.

The Fourier expansion has the nice property that when a series is
truncated at $N$, the remaining terms still give the best Fourier
expansion of the function up to $N$. In other words, the
coefficients of a truncated Fourier series are equal to the first
$N$ coefficients of a full Fourier series.

As an example, in what follows we use the pseudo-Fourier method to
simulate a pattern generated by the following test function
\begin{equation}
\label{23} F(\phi)=|\sin\phi|,
\end{equation}
which can be expanded as a Fourier series
\begin{equation}
\label{24} F(\phi)=\frac{2}{\pi}-\frac{4}{\pi}\sum_{n=1}^{\infty}
\frac{ \cos (2n\phi)}{(4n^2-1)},
\end{equation}

Comparing Eq. (\ref{24}) with (\ref{20}) and using (\ref{22}) we
find the subsidiary phase $\theta_m$ and the superposition
coefficients $C_n$ in Eqs. (\ref{10}) and (\ref{15}) to be
\begin{eqnarray}
\label{25} \theta_m&=&m\pi, \hspace{0.5cm} C_{2n+1}=0,
\\
\label{26} |C_{2n}|^2&=& \frac{4}{\pi t\sin(2\gamma)(4n^2-1)}.
\end{eqnarray}

Substituting Eqs. (\ref{25}) and (\ref{26}) into Eq. (\ref{15}),
one can obtain the superposition state to realize the pattern of
the test function given by Eq. (\ref{23}). And from (\ref{25}),
(\ref{26}) and (\ref{15})  we can see that the superposition state
consists of even-number-photon  NMES like (\ref{10}), and only the
module of the superposition coefficients affect the deposition
rate of quantum photolithography. In Figure 1 we have simulated
the test pattern (\ref{23}) by using the superposition state given
by Eq. (\ref{22}) for $N=2$, $6$, and $12$ cases, respectively.
From Figure 1 we can see that the calculated patterns are in good
agreement with the test pattern as shown by the solid curve, and
the larger the value of $N$ is, the better the effect of the
simulation.

\section{ Concluding remarks}
In conclusion we have generalized the $N$-photon MES quantum
photolithography to the $N$-photon NMES case, and  investigated
the correlations between quantum entanglement and quantum
photolithography. It has been found that quantum photolithography
can be manipulated  and controlled through varying quantum
entanglement between two applied light beams. Especially, for the
nonlocal entanglement case, we have showed that  the amplitude of
the deposition function increases with the degree of entanglement.
Hence one can manipulate and control the amplitude of the
deposition rate through varying the quantum entanglement between
the two light beams while the resolution of quantum lithography
remains unchanged. And for the local entanglement case, we have
found that the local entanglement between the two light beams can
enhance the effective Rayleigh resolution of quantum
photolithography. However, it would be challenging to create
locally entangled states.

\acknowledgments This work is supported by the National
Fundamental Research Program Grant No. 2001CB309310, the National
Natural Science Foundation Grant Nos. 90203018 and 10075018, the
State Education Ministry of China, the Educational Committee of
Hunan Province, and the Innovation Funds from Chinese Academy of
Sciences via the Institute of Theoretical Physics, Academia,
Sinica.


\begin{references}
\bibitem{r1}  Rathe U V and  Scully M O, 1995 {\it Lett. Math. Phys.} {\bf 34} 297
\bibitem{r2}  D'Angelo M,  Chekhova M V, and  Shih Y 2001 {\it Phys. Rev. Lett.} {\bf 87} 013602
\bibitem{r3}  Boto A N,  Kok P,  Abrams D S,  Braunstein S L,
             Williams C P, and  Dowling J P 2000 {\it Phys. Rev. Lett.} {\bf 85}, 2733
\bibitem{r3'}  Lugiato L A, Gatti A and  Brambilla E 2002 {\it J. Opt. B: Quantum Semiclass. Opt.} {\bf 4} S1
\bibitem{r4}  Br\"{u}ck S R J {\it et al.}, 1998 {\it Microelectron. Eng.} {\bf 42} 145
\bibitem{r5} Mack C A 1996 {\it  Opt. Photonics News } {\bf 7} 29
\bibitem{r6} Manuscripure M and  Liang R 2000 {\it  Opt. Photonics News} {\bf  11} 36
\bibitem{r7} Kok P,  Boto A N, Abrams D S,  Williams C P, Braunstein S L, and Dowling J P 2001 {\it Phys. Rev.} {\bf A
63} 063407
\bibitem{r8} Bj\"{o}rk G,  Sanchez-Soto L L and  S\"{a}derholm J 2001 {\it Phys. Rev.} {\bf A 64} 013811
\bibitem{r9} Bj\"{o}rk G,  Sanchez-Soto L L and  S\"{a}derholm J 2001 {\it Phys. Rev. Lett.} {\bf 86} 4516
\bibitem{r10} Rayleigh L 1879 {\it Philos. Mag.} {\bf 8} 261;
              Born M  and Wolf E 1980  {\it Principles of Optics} (Pergamon Press. New York. 1980), 6th ed., Sec. 7.6.3.
\bibitem{r11} Narozhny N B 1981 {\it Phys. Rev.} {\bf  A 23} 236
\bibitem{r12} Rempe G, Walther H and  Klein N 1987 {\it  Phys. Rev. Lett.} {\bf  58} 353
\end{references}
\end{document}